\documentclass[preprint,superscriptaddress,showpacs,tightenlines]{revtex4}
\usepackage{amssymb}
\usepackage{amsmath}
\usepackage{graphicx,bm}

\input{tcilatex}

\begin{document}

\title{Influence of a single defect on the conductance of a tunnel point
contact between a normal metal and a superconductor}
\author{Ye.S. Avotina}
\affiliation{B.I. Verkin Institute for Low Temperature Physics and Engineering, National
Academy of Sciences of Ukraine, 47, Lenin Ave., 61103, Kharkov,Ukraine.}
\affiliation{Kamerlingh Onnes Laboratorium, Universiteit Leiden, Postbus 9504, 2300
Leiden, The Netherlands.}
\author{Yu.A. Kolesnichenko}
\affiliation{B.I. Verkin Institute for Low Temperature Physics and Engineering, National
Academy of Sciences of Ukraine, 47, Lenin Ave., 61103, Kharkov,Ukraine.}
\affiliation{Kamerlingh Onnes Laboratorium, Universiteit Leiden, Postbus 9504, 2300
Leiden, The Netherlands.}
\author{J.M. van Ruitenbeek}
\affiliation{Kamerlingh Onnes Laboratorium, Universiteit Leiden, Postbus 9504, 2300
Leiden, The Netherlands.}

\begin{abstract}
We have investigated theoretically the conductance of a
Normal-Superconductor point-contact in the tunnel limit and analyzed the
quantum interference effects originating from the scattering of
quasiparticles by point-like defects. Analytical expressions for the
oscillatory dependence of the conductance on the position of the defect are
obtained for the defect situated either in the normal metal, or in the
superconductor. It is found that the amplitude of oscillations significantly
increases when the applied bias approaches the gap energy of the
superconductor. The spatial distribution of the order parameter near the
surface in the presence of a defect is also obtained.
\end{abstract}

\maketitle

\section{\protect\bigskip Introduction}

Electron scattering by single surface \cite{Cromme} and subsurface \cite%
{babble} defects results in an oscillatory dependence of the Scanning
Tunnelling Microscope (STM) conductance $G$ on the distance, $r_{0}$,
between the contact and the defect. These oscillations originate from the
interference of electron waves, which are scattered by the defect and
reflected back by the contact. They have the same period ($G\sim \sin \left(
2k_{F}r_{0}+\delta \right) $, $k_{F}$ is the Fermi wave vector) as the
Friedel oscillations \cite{Friedel} of the local electron density of states
in the vicinity of a scatterer. For subsurface point-like defects the
oscillatory dependence of the conductance in a STM-like geometry has been
investigated theoretically in Refs.~\cite%
{Avotina1,Avotina2,Avotina2a,Avotina3,Avotina4}.

Although defects below a metal surface can be 'visible' in STM data for up
to ten interatomic distances \cite{Wend,Quaas}, the amplitude of the quantum
oscillations in the conductance become very small with increasing defect
depth. An effective way to enhance the STM sensitivity to such oscillation
effects is to use a superconducting tip \cite{Pan}. In Ref.~\cite{Stwave}
using a low-temperature STM with normal metal tungsten tips and
superconducting niobium tips, the formation of electron standing waves near
surface defects and step edges on a Au (111) surface have been observed. It
was demonstrated that the amplitude of conductance oscillations is
significantly enhanced when a superconducting tip is used, and when the
applied bias $\left\vert eV\right\vert $ is close to the gap energy $\Delta
_{0}$ of the superconductor.

The investigation of various defects in superconductors with STM is of
interest by itself. For example, in Ref.~\cite{MagImp} a bound state near a
magnetic Mn adatom on the surface of superconducting Nb was observed by STM.
The effect of single Zn defects on the superconductivity in high-T$_{\mathrm{%
c}} $ superconductors was investigated in Ref.~ \cite{Zn}, and the
manifestation of d-wave symmetry of the order parameter was observed in the
quasibound state near the defect.

The listed reasons define the interest of theoretical investigations on the
conductance of normal metal - superconductor (NS) tunnel contacts of small
lateral size, in the vicinity of which a single defect is placed. The
authors of Ref.~\cite{Prada} considered the conductance of a NS contact of
finite size at low temperatures and for voltages $\left\vert eV\right\vert
<\Delta _{0}$ using the tunnelling Hamiltonian approximation. They found
that, when the radius $a$ of the contact is smaller then the Fermi wave
length $\lambda _{F},$ the conductance of a NS point-contact becomes $%
G_{ns}=\left( h/2e^{2}\right) G_{nn}^{2}\sim a^{8}$, where $G_{nn}$ is the
conductance of the contact in the normal state \cite{Prada}. This dependence
is fundamentally different from the result of a quasiclassical theory \cite%
{Zaitsev}, valid for $a\gg $ $\lambda _{F}.$

The conductivity of large ($a\gg $ $\lambda _{F}$) ballistic NS contacts in
the presence of a `planar defect' was investigated theoretically in several
papers \cite{Hahn,Son,Bagwell1,Bagwell2}. In these papers a planar NS
structure and a $\delta$-functional potential barrier, playing the role of
the defect, have been considered, from which `geometrical' resonances
resulted due to combined Andreev and normal reflections.

In order to describe the effect of isolated point-like defects in a
superconductor on the STM conductance usually calculations of the local
density of states $n(\mathbf{r})$ are used (for a review, see \cite{Balatsky}%
), where it is assumed that the conductance of the small tunnel contact is
proportional to the local density of electron states. While for subsurface
defects this assumption remains qualitatively valid, it does not permit a
correct description of the details of the conductance oscillations because
the bulk electron density of states around the defect is modified by
reflection from the interface, $\mathbf{r\in \Sigma }$, and in the limit of
zero tunnelling probability we have $n\left( \mathbf{r\in \Sigma }\right) =0.
$ In this case, the problem of electron transmission through the small NS
tunnel junction in the presence of the defect should be considered.

In this paper we present the results of a theoretical investigation of the
conductance of a NS point contact (with $a \ll \lambda _{F}$ ) in the
tunnelling limit and we analyze the quantum interference effects originating
from the scattering of quasiparticles by a point-like defect. Analytical
expressions are obtained for the dependence of the conductance on the
position of the defect and on the applied voltage, for the defect situated
in the normal metal or in the superconductor.

\section{Model and basic equations}

Our model is presented in the Fig.1. The normal and superconducting
half-spaces are separated by an infinitely thin dielectric interface, which
has an orifice of radius $a.$ The potential barrier in the plane of
interface $z=0$ is taken to be a $\delta -$function, $U\left( \mathbf{r}%
\right) =U_{0}f\left( \rho \right) \delta \left( z\right) ,$ where $\rho $
is the value of the radius vector $\mathbf{\rho }$ in the plane $z=0$. The
function $f\left( \rho \right) \rightarrow \infty $ in all points of the
plane except in the contact ($\rho <a$) , where $f\left( \rho \right) =1$.
In the point $\mathbf{r}_{0}$ a nonmagnetic defect described by a
spherically symmetric potential $D\left( \left\vert \mathbf{r-r}%
_{0}\right\vert \right) $ is placed. A voltage $V$ is applied between the
two sides of the contact. We assume that the transmission probability $%
\left\vert t\right\vert $ of electrons through the barrier in the orifice is
small ($\left\vert t\right\vert \approx \hbar ^{2}k_{F}/m^{\ast }U_{0}\ll 1$%
, $m^{\ast }$ is effective electron mass). In that case the applied voltage
drops entirely over the barrier and the electric potential can be described
by a step function, $V\left( z\right) =V\,\Theta \left( -z\right) $ with $V$
a constant. Based on the same reasoning we use a step function for the
superconducting order parameter $\Delta \left( \mathbf{r}\right) =\Delta
\left( \mathbf{r}\right) \Theta \left( z\right) $. We consider the case of
low temperatures and in the calculations take $T=0.$ At zero temperature a
tunnel current flows through the contact for $\left\vert eV\right\vert
>\Delta $ $.$ The applied bias is assumed to be small on the scale of the
Debye frequency $\omega _{D}$ and the Fermi energy $\varepsilon _{F}$, $%
\left\vert eV\right\vert \ll \hbar \omega _{D}\ll \varepsilon _{F}$. 
\begin{figure}[tbp]
\includegraphics[width=10cm,angle=0]{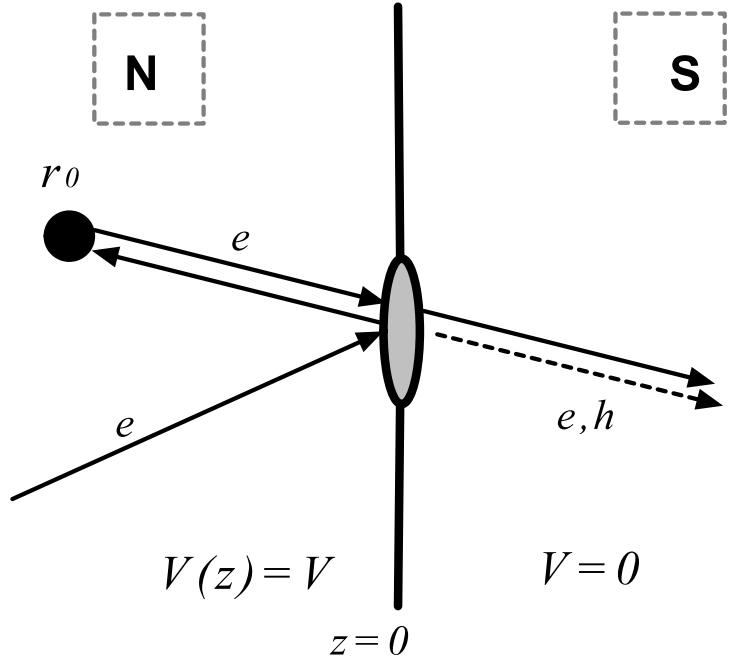}
\caption{Model of the contact. The point-like defect is situated in the
normal half-space. The electron trajectories in the normal metal and the
trajectories of `electron-like' and `hole-like' excitations in the
superconductor are shown schematically.}
\end{figure}

For definiteness we consider electron tunnelling from the normal half-space $%
\left( z<0\right) $ to the superconducting half-space $\left( z>0\right) $,
i.e. $eV>0.$ In order to evaluate the total current through the contact, $%
I\left( V\right) $, and the differential conductance, $G\left( V\right)
=dI\left( V\right) /dV$, we should find the current density $\mathbf{j}_{%
\mathbf{k}}\left( \mathbf{r}\right) $ of quasiparticles with momentum $%
\mathbf{k}$ at $z>0,$ formed by electrons transmitted through the contact.
The current density $\mathbf{j}_{\mathbf{k}}\left( \mathbf{r}\right) $ can
be expressed in terms of the coefficients $u_{\mathbf{k}}\left( \mathbf{r}%
\right) $ and $v_{\mathbf{k}}\left( \mathbf{r}\right) $ of the canonical
Bogoliubov transformation \cite{Svidz,Hard} 
\begin{equation}
\mathbf{j}_{\mathbf{k}}\left( \mathbf{r}\right) =\frac{e\hbar }{m^{\ast }}%
\func{Im}\left[ u_{\mathbf{k}}(\mathbf{r})\nabla u_{\mathbf{k}}^{\ast }(%
\mathbf{r})f_{\mathrm{F}}\left( E_{\mathbf{k}}\right) -v_{\mathbf{k}}(%
\mathbf{r})\nabla v_{\mathbf{k}}^{\ast }(\mathbf{r})f_{\mathrm{F}}\left( -E_{%
\mathbf{k}}\right) \right] ,  \label{j_k(r)}
\end{equation}%
where $f_{\mathrm{F}}\left( E\right) $ is the Fermi function, which at $T=0$
is simply the unit step-function, $f_{\mathrm{F}}\left( E\right) =\Theta
\left( E\right) .$ The functions $u_{\mathbf{k}}\left( \mathbf{r}\right) $
and $v_{\mathbf{k}}\left( \mathbf{r}\right) $ satisfy to the Bogoliubov-de
Gennes (BdG) equations \cite{Gennes} 
\begin{eqnarray}
\left[ -\frac{\hbar ^{2}}{2m^{\ast }}\nabla ^{2}-\varepsilon _{F}+D\left(
\left\vert \mathbf{r-r}_{0}\right\vert \right) \right] u_{\mathbf{k}}\left( 
\mathbf{r}\right) +\Delta \left( \mathbf{r}\right) v_{\mathbf{k}}\left( 
\mathbf{r}\right)  &=&E_{\mathbf{k}}u_{\mathbf{k}}\left( \mathbf{r}\right) ,
\label{BdG} \\
-\left[ -\frac{\hbar ^{2}}{2m^{\ast }}\nabla ^{2}-\varepsilon _{F}+D\left(
\left\vert \mathbf{r-r}_{0}\right\vert \right) \right] v_{\mathbf{k}}\left( 
\mathbf{r}\right) +\Delta ^{\ast }\left( \mathbf{r}\right) u_{\mathbf{k}%
}\left( \mathbf{r}\right)  &=&E_{\mathbf{k}}v_{\mathbf{k}}\left( \mathbf{r}%
\right) .  \notag
\end{eqnarray}%
Eqs.~(\ref{BdG}) may be interpreted as wave equations for a two-component
`wave function', 
\begin{equation}
\widehat{\psi }_{\mathbf{k}}=\left( 
\begin{array}{c}
u_{\mathbf{k}} \\ 
v_{\mathbf{k}}%
\end{array}%
\right) ,  \label{psi}
\end{equation}%
of quasiparticles with energy $E_{\mathbf{k}}.$ The conditions, which
connect the vector $\widehat{\psi }_{\mathbf{k}}$ in the normal metal $(%
\widehat{\psi }_{n\mathbf{k}})$ and in the superconductor $(\widehat{\psi }%
_{s\mathbf{k}})$ at the interface $z=0$ are 
\begin{equation}
\widehat{\psi }_{n\mathbf{k}}\left( \rho ,0\right) =\widehat{\psi }_{s%
\mathbf{k}}\left( \rho ,0\right) =\widehat{\psi }_{\mathbf{k}}\left( \rho
,0\right)   \label{continue}
\end{equation}%
\begin{equation}
\frac{\partial }{\partial z}\widehat{\psi }_{s\mathbf{k}}\left( \rho
,0\right) -\frac{\partial }{\partial z}\widehat{\psi }_{n\mathbf{k}}\left(
\rho ,0\right) =\frac{2m^{\ast }}{\hbar ^{2}}U_{0}f\left( \rho \right) 
\widehat{\psi }_{\mathbf{k}}\left( \rho ,0\right)   \label{deriv}
\end{equation}%
\textbf{\ } The order parameter in the superconductor should be determined
from the self-consistently condition%
\begin{equation}
\Delta \left( \mathbf{r}\right) =\gamma \sum\limits_{\mathbf{k,}E_{\mathbf{k}%
}<\hbar \omega _{D}}u_{\mathbf{k}}\left( \mathbf{r}\right) v_{\mathbf{k}%
}^{\ast }\left( \mathbf{r}\right) \left[ 1-2f_{\mathrm{F}}\left( E_{\mathbf{k%
}}\right) \right] ,  \label{delta}
\end{equation}%
\begin{equation}
\Delta \left( z\rightarrow +\infty \right) \rightarrow \Delta _{0},
\label{delta_inf}
\end{equation}%
where the constant $\Delta _{0}$ can be chosen real; $\gamma $ is the pair
potential constant. It can be easily shown \cite{Svidz} that Eq.~(\ref%
{j_k(r)}) combined with the self-consistently condition (\ref{delta})
automatically satisfies to the continuity equation%
\begin{equation}
\text{div}\sum\limits_{\mathbf{k}}\mathbf{j}_{\mathbf{k}}\left( \mathbf{r}%
\right) =0.  \label{divj}
\end{equation}

The current-voltage characteristic $I\left( V\right) $ of the contact in the
presence of a defect can be found by means of integration of the current
density $\mathbf{j}_{\mathbf{k}}\left( \mathbf{r}\right) $ over the momentum 
$\mathbf{k}$ (within the energy interval $\Delta _{0}\leq E_{\mathbf{k}}\leq
eV$ ) and over a surface overlapping the contact in the superconducting
half-space. For this surface we choose a half-sphere of large radius $r\gg
r_{0},\xi _{0}$ ($\xi _{0}$ is the coherence length of the superconductor)
centered at the contact $r=0.$ On this half-sphere we assume $\Delta \left( 
\mathbf{r}\right) =\Delta _{0}$ and hence $E_{\mathbf{k}}=\sqrt{\xi _{%
\mathbf{k}}^{2}+\Delta _{0}^{2}},$ where $\xi _{\mathbf{k}}=\hbar
^{2}k^{2}/2m^{\ast }-\varepsilon _{F}$ is the kinetic energy measured from
the Fermi level. The conductance $G\left( V\right) $ of the contact (at $T=0$%
) is given by 
\begin{equation}
G\left( V\right) =4\pi r\,e^{2}N\left( 0\right) \int \frac{d\Omega }{4\pi }%
\Theta \left( z\right) \int\limits_{-\infty }^{\infty }d\xi _{\mathbf{k}%
}\int \frac{d\Omega _{\mathbf{k}}}{4\pi }\Theta \left( k_{z}\right) \left( 
\mathbf{rj}_{\mathbf{k}}\left( \mathbf{r}\right) \right) \delta \left( E_{%
\mathbf{k}}-eV\right) .  \label{G(V)}
\end{equation}%
\textbf{\ } where $d\Omega $ and $d\Omega _{\mathbf{k}}$ are elements of
solid angle in the real and momentum spaces, respectively, $N\left( 0\right) 
$ is the density of states for one direction of spin.

\section{Solution of the Bogoliubov - de Gennes equation}

Generally, a self-consistent solution of Eqs. (\ref{BdG}) can be found only
numerically. Such solution must fulfil the condition of conservation of the
total current $I$ through any surface overlapping the contact, in spite of
the spatial dependence of the order parameter. In order to simplify the task
we will exploit the condition of a small barrier transparency and find an
analytical solution of Eqs. (\ref{BdG}) using the approximation of a
constant order parameter $\Delta \left( \mathbf{r}\right) =\Delta _{0}\Theta
\left( z\right) .$ By means of this solution the coordinate dependence of $%
\Delta \left( \mathbf{r}\right) $ can be found (see Appendix).

In this section we generalize the method developed in the papers \cite%
{KMO,Avotina1}. We search the solutions of Eqs. (\ref{BdG}) as an expansion
into a series over the small transmission probability $\left\vert
t\right\vert \sim 1/U_{0}$, 
\begin{equation}
\widehat{\psi }_{\mathbf{k}}\left( \mathbf{r}\right) =\widehat{\psi }_{%
\mathbf{k}0}\left( \mathbf{r}\right) +\widehat{\psi }_{\mathbf{k}1}\left( 
\mathbf{r}\right) +\dots ,  \label{expand}
\end{equation}%
where $\widehat{\psi }_{\mathbf{k}0}\left( \mathbf{r}\right) $ satisfies the
zero-boundary condition at $z=0,$ and $\widehat{\psi }_{\mathbf{k}1}\left( 
\mathbf{r}\right) \sim 1/U_{0}.$ For the calculation of the current in
leading approximation in the transmission coefficient $\left( I\sim
1/U_{0}^{2}\right) $ it is enough to find the first correction $\widehat{%
\psi }_{\mathbf{k}1}\left( \mathbf{r}\right)$. Substituting the expansion (%
\ref{expand}) into the boundary conditions (\ref{continue}), (\ref{deriv})
we find that the function $\widehat{\psi }_{\mathbf{k}1}\left( \mathbf{r}%
\right) $ satisfies the condition of continuity at $z=0$, and its value at $%
z=+0$ (in the superconducting half-space) is given by the relations 
\begin{equation}
u_{s\mathbf{k}1}\left( \rho ,0\right) =\frac{\hbar ^{2}}{2m^{\ast
}U_{0}f\left( \rho \right) }\frac{\partial }{\partial z}u_{n\mathbf{k}%
0}\left( \rho ,0\right) ;\quad v_{\mathbf{sk}1}\left( \rho ,0\right) =0.
\label{bound_condition}
\end{equation}%
The boundary condition does not contain Andreev reflections, which appear in
the next approximation in $1/U_{0}$ \cite{BTK}. Thus, we will not consider
Andreev resonances, which were analyzed in Refs. \cite%
{Hahn,Son,Bagwell1,Bagwell2} for a one-dimensional model.

The quasiparticle scattering by the defect will be taken into account by
perturbation theory in the strength of the interaction with the defect.
First, we find the solution of Eqs. (\ref{BdG}) for the contact without
defect.

Let us consider an electron with energy $E_{\mathbf{k}}>\Delta_{0}$, which
moves towards the interface from the normal metal. When $D\left( \mathbf{r}
\right) =0$ (the defect is absent) and $1/U_{0}=0$ (the interface is
impenetrable for electrons), in the normal half-space we have 
\begin{equation}
u_{n\mathbf{k}0}\left( \mathbf{r}\right) =e^{i\mathbf{\varkappa \rho }%
}\left( e^{ik_{z}z}-e^{-ik_{z}z}\right) ,\quad v_{n\mathbf{k}0}\left( 
\mathbf{r}\right) =0,  \label{uv0_normal}
\end{equation}%
where $\mathbf{k=}\left( \mathbf{\varkappa ,}k_{z}\right) ,$ $k_{z}=k\cos
(\vartheta) ,$ $\vartheta $ is the angle between the vector $\mathbf{k}$ and
the $z$ axis, and $\mathbf{\varkappa }$ is the component of the wave vector
parallel to the interface.

Making use of the Fourier transform of the $\widehat{\psi }_{\mathbf{k}%
}\left( \mathbf{r}\right) $ components over the coordinate $\mathbf{\rho }$
in the plane parallel to the interface, 
\begin{equation}
\widehat{\psi }_{\mathbf{k}1}\left( \mathbf{\rho },z\right)
=\int\limits_{-\infty }^{\infty }d\mathbf{\varkappa }^{\prime }\widehat{\Psi 
}_{\mathbf{k}1}\left( \mathbf{\varkappa }^{\prime },z\right) e^{i\mathbf{%
\varkappa }^{\prime }\mathbf{\rho }},
\end{equation}%
and finding $\widehat{\Psi }_{\mathbf{k}1}\left( \mathbf{\varkappa }^{\prime
},0\right) $ from the simplified boundary condition (\ref{bound_condition}),
we find the solution of Eqs. (\ref{BdG}) in the superconducting half-space 
\begin{equation}
u_{\mathbf{k}1}\left( \mathbf{r}\right) =t\left( k_{z}\right) \frac{1}{%
u_{0}^{2}-v_{0}^{2}}\left[ u_{0}^{2}\varphi _{0}^{\left( +\right) }\left( 
\mathbf{r}\right) +v_{0}^{2}\varphi _{0}^{\left( -\right) }\left( \mathbf{r}%
\right) \right] ,  \label{u0}
\end{equation}%
\begin{equation}
v_{\mathbf{k}1}\left( \mathbf{r}\right) =t\left( k_{z}\right) \frac{%
u_{0}v_{0}}{u_{0}^{2}-v_{0}^{2}}\left[ \varphi _{0}^{\left( +\right) }\left( 
\mathbf{r}\right) +\varphi _{0}^{\left( -\right) }\left( \mathbf{r}\right) %
\right] ,  \label{v0}
\end{equation}%
where%
\begin{equation}
\varphi _{0}^{\left( \pm \right) }\left( \mathbf{r}\right) =\pm \frac{1}{%
\left( 2\pi \right) ^{2}}\int\limits_{-\infty }^{\infty }d\mathbf{\varkappa }%
^{\prime }e^{i\mathbf{\varkappa }^{\prime }\mathbf{\rho }}\int\limits_{-%
\infty }^{\infty }d\mathbf{\rho }^{\prime }\frac{e^{i\left( \mathbf{%
\varkappa -\varkappa }^{\prime }\right) \mathbf{\rho }^{\prime }}}{f\left(
\rho \right) }e^{\pm ik_{z}^{\left( \pm \right) }z},  \label{fi_0}
\end{equation}%
\begin{equation}
k_{z}^{\left( \pm \right) }=\frac{\sqrt{2m^{\ast }}}{\hbar }\left[
\varepsilon _{F}-\frac{\hbar ^{2}\varkappa ^{2}}{2m^{\ast }}\pm \sqrt{E_{%
\mathbf{k}}^{2}-\Delta _{0}^{2}}\right] ^{1/2},  \label{k_z}
\end{equation}%
\begin{equation}
u_{0}^{2}=1-v_{0}^{2}=\frac{1}{2}\left( 1+\frac{\xi _{\mathbf{k}}}{E_{%
\mathbf{k}}}\right) 
\end{equation}%
$t\left( k_{z}\right) =\hbar ^{2}k_{z}/im^{\ast }U_{0}$ is the amplitude of
electron wave after tunnelling through the homogeneous barrier with a large $%
U_{0}.$ The functions $u_{\mathbf{k}1}\left( \mathbf{r}\right) $ and $v_{%
\mathbf{k}1}\left( \mathbf{r}\right) $ contain the sum of two solutions $%
\varphi _{0}^{\left( \pm \right) }\left( \mathbf{r}\right) $ of Eqs.~(\ref%
{BdG}), which correspond to `electron-like' $(k_{z}^{\left( +\right)
}>k_{zF}=\frac{1}{\hbar }\sqrt{2m^{\ast }\left( \varepsilon _{F}-\hbar
^{2}\varkappa ^{2}/2m^{\ast }\right) })$ and `hole-like' $(k_{z}^{\left(
-\right) }<k_{zF})$ quasiparticles having a positive $z$-component of the
group velocity $\mathbf{v}_{g}=dE_{\mathbf{k}}/\hbar d\mathbf{k}$. 

For a small radius of the contact (in the limit $a\rightarrow 0$) the
function (\ref{fi_0}) takes the form \cite{Avotina4}%
\begin{equation}
\varphi _{0}^{\left( \pm \right) }\left( \mathbf{r},k\right) =\frac{\left(
k^{\left( \pm \right) }a\right) ^{2}\cos \theta }{2}h_{1}^{\left( 1\right)
}( k^{\left( \pm \right) }r) ,
\end{equation}%
\begin{equation}
k^{\pm }( E_{\mathbf{k}}) =\frac{\sqrt{2m^{\ast }}}{\hbar }\left[
\varepsilon _{F}\pm \sqrt{E_{\mathbf{k}}^{2}-\Delta _{0}^{2}}\right] ^{1/2}.
\end{equation}%
Here, $h_{1}^{\left( 1\right) }( x) $ is the spherical Hankel function of
the first kind.

In the presence of the defect the functions $u_{\mathbf{k}1}\left( \mathbf{r}%
\right) $ and $v_{\mathbf{k}1}\left( \mathbf{r}\right) $ can be found in
first approximation in the potential $D\left( \left\vert \mathbf{r-r}%
_{0}\right\vert \right) $ of electron-impurity interaction by means of the
Eqs. (\ref{BdG}).

1) If the defect is situated in the normal half-space the functions $u_{%
\mathbf{k}1}\left( \mathbf{r}\right) $ and $v_{\mathbf{k}1}\left( \mathbf{r}%
\right) $ in the superconductor have the same form as Eqs. (\ref{u0}), (\ref%
{v0}) in which the amplitude $t\left( k_{z}\right) $ must be replaced by the
value 
\begin{equation}
\widetilde{t}\left( k_{z}\right) =t\left( k_{z}\right) +\frac{4\pi
^{2}m^{\ast }k}{\hbar ^{2}}gt\left( k\right) u_{n\mathbf{k}0}\left( \mathbf{r%
}_{0}\right) h_{1}^{\left( 1\right) }\left( kr_{0}\right) ,  \label{t1}
\end{equation}%
where $g$ is the constant of the electron interaction with the defect%
\begin{equation}
g=\int d\mathbf{r}D\left( \left\vert \mathbf{r-r}_{0}\right\vert \right) .
\label{g}
\end{equation}%
In order to obtain Eq.(\ref{t1}) we assume that the characteristic radius of
the scattering potential is much smaller than the Fermi wave length $\lambda
_{F}$ (point defect). This condition permits taking the functions $u_{n%
\mathbf{k}0}( \mathbf{r}) $ and $h_{1}^{( 1) }( kr) $ outside the integral
at the point $\mathbf{r}=\mathbf{r}_{0}$. The variations in the amplitudes
of the `wave functions' $u_{\mathbf{k}1}\left( \mathbf{r}\right) $ and $v_{%
\mathbf{k}1}\left( \mathbf{r}\right) $ result from the fact that the wave
incident to the contact is a superposition of a plane wave and a spherical
wave that comes from the scattering by the defect.

2) If the defect is situated inside the superconductor, the additions $%
\Delta u_{\mathbf{k}1}\left( \mathbf{r}\right) $ and $\Delta v_{\mathbf{k}%
1}\left( \mathbf{r}\right) $ to the functions (\ref{u0}), (\ref{v0}) due to
the defect scattering take the form%
\begin{eqnarray}
\Delta u_{\mathbf{k}1}\left( \mathbf{r}\right) &=&\frac{2\pi m^{\ast }g}{%
\hbar ^{2}}\frac{1}{v_{0}^{2}-u_{0}^{2}}\int\limits_{-\infty }^{\infty }d%
\mathbf{\varkappa }e^{i\mathbf{\varkappa }\left( \mathbf{\rho }-\mathbf{\rho 
}_{0}\right) }\left\{ \frac{1}{k_{z}^{\left( +\right) }}u_{0}\sin \left(
k_{z}^{\left( +\right) }z\right) e^{ik_{z}^{\left( +\right) }z}\left[
u_{0}u_{\mathbf{k}1}\left( \mathbf{r}_{0}\right) -v_{0}v_{\mathbf{k}1}\left( 
\mathbf{r}_{0}\right) \right] \right.  \notag \\
&&\left. +\frac{1}{k_{z}^{\left( -\right) }}v_{0}\sin \left( k_{z}^{\left(
-\right) }z\right) e^{-ik_{z}^{\left( -\right) }z}\left[ u_{0}v_{\mathbf{k}%
1}\left( \mathbf{r}_{0}\right) -v_{0}u_{\mathbf{k}1}\left( \mathbf{r}%
_{0}\right) \right] \right\} ;
\end{eqnarray}%
\begin{eqnarray}
\Delta v_{\mathbf{k}1}\left( \mathbf{r}\right) &=&\frac{2\pi m^{\ast }g}{%
\hbar ^{2}}\frac{1}{v_{0}^{2}-u_{0}^{2}}\int\limits_{-\infty }^{\infty }d%
\mathbf{\varkappa }e^{i\mathbf{\varkappa }\left( \mathbf{\rho }-\mathbf{\rho 
}_{0}\right) }\left\{ \frac{1}{k_{z}^{\left( +\right) }}v_{0}\sin \left(
k_{z}^{\left( +\right) }z\right) e^{ik_{z}^{\left( +\right) }z}\left[
u_{0}u_{\mathbf{k}1}\left( \mathbf{r}_{0}\right) -v_{0}v_{\mathbf{k}1}\left( 
\mathbf{r}_{0}\right) \right] \right.  \notag \\
&&\left. -\frac{1}{k_{z}^{\left( -\right) }}u_{0}\sin \left( k_{z}^{\left(
-\right) }z\right) e^{-ik_{z}^{\left( -\right) }z}\left[ u_{0}v_{\mathbf{k}%
1}\left( \mathbf{r}_{0}\right) -v_{0}u_{\mathbf{k}1}\left( \mathbf{r}%
_{0}\right) \right] \right\} .
\end{eqnarray}%
\qquad \qquad \qquad

It is known that the order parameter $\Delta \left( \mathbf{r}\right) $
displays Friedel-like oscillations near a defect \cite{Fetter,Flatte} or a
surface \cite{Boyd,Tanaka}. The current through the tunnel contact $I$ is
defined by the average value of $\Delta \left( \mathbf{r}\right) $, which
coincides with $\Delta _{0}.$ In the Appendix we analyze the spatial
dependence of $\Delta \left( \mathbf{r}\right) $ near the surface of the
superconductor, in the vicinity of which a non-magnetic defect is placed (at
the distance less than the coherence length $\xi _{0}$). Figure 2
illustrates the results of these calculations. An inhomogeneous spatial
distribution of the order parameter is visible. We removed from the plot the
region of radius $\lambdabar _{F}$ (black circle) near the defect where Eq.~(%
\ref{D(r)}) is not valid.

\begin{figure}[tbp]
\includegraphics[width=10cm,angle=0]{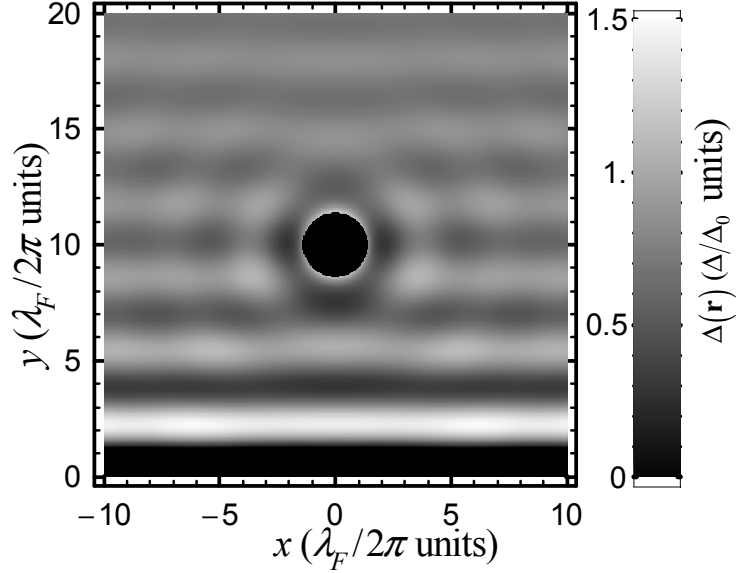}
\caption{Real space image of $\Delta \left( \mathbf{r}\right) /\Delta _{0}$
near the surface of the superconductor in the plane passing through the
defect which has been obtained by using Eq.~(\protect\ref{D(r)}), and the
parameters $z_{0}=10\lambdabar _{F},$ $\protect\xi _{0}=10^{4}$ $\lambdabar
_{F},$ $\widetilde{g}=4\protect\pi $.}
\end{figure}

\section{Conductance of the contact}

By means of the solutions of the BdG equations, which have been obtained in
previous section, we calculated the conductance $G$ of the NS tunnel point
contact. In linear approximation in the electron-defect interaction constant 
$g$ the conductance $G$ can be presented as the sum of two terms, 
\begin{equation}
G\left( V,r_{0}\right) =G_{0ns}\left( V\right) +\Delta G_{osc}\left(
V,r_{0}\right) ,\quad eV>\Delta _{0}.  \label{G}
\end{equation}%
The first term, $G_{0ns}\left( V\right) $, in Eq.~(\ref{G}) is the
conductance of the NS tunnel point contact in the absence of the defect%
\begin{equation}
G_{0ns}\left( V\right) =G_{0nn}\frac{eV}{\sqrt{\left( eV\right) ^{2}-\Delta
_{0}^{2}}};\quad G_{0nn}=\frac{2e^{2}a^{4}m^{\ast }\varepsilon _{F}^{3}}{%
9\pi \hbar ^{3}U_{0}^{2}},  \label{G0}
\end{equation}%
where $G_{0nn}$ is the conductance of a contact between normal metals, which
is multiplied by the normalized density of states of the superconductor at $%
E=eV$ in Eq.~(\ref{G0}). The second term describes the oscillatory
dependence of the conductance on the distance between the contact and the
defect.

If the defect is situated in the normal metal half-space $\Delta
G_{osc}\left( V,r_{0}\right) $ is given by%
\begin{equation}
\Delta G_{osc}\left( V,r_{0}\right) =-G_{0ns}\left( V\right) \frac{12}{\pi }%
\widetilde{g}\left( \frac{\lambdabar _{F}}{r_{0}}\right) ^{2}\left(
k_{F}z_{0}\right) ^{2}j_{1}\left( k_{F}r_{0}\right) y_{1}\left(
k_{F}r_{0}\right) ,  \label{dG_N}
\end{equation}%
where 
\begin{equation}
\widetilde{g}=\frac{2\pi m^{\ast }k_{F}}{\hbar ^{2}}g
\end{equation}%
is the dimensionless electron-defect interaction constant, $j_{l}(x)$ and $%
y_{l}(x)$ are the spherical Bessel functions of the first and the second
kind \cite{AS}, and $\lambdabar _{F}=\hbar /\sqrt{2m^{\ast }\varepsilon _{F}}
$ . In Fig.3 dependencies of $\Delta G_{osc}\left( V,r_{0}\right) $ on the
distance $\rho _{0}$ are shown for two values of the bias $eV,$ one of which
is very close to the gap energy ($eV/\Delta _{0}=1.1),$ and the second one
is $eV=2\Delta _{0}.$ The figure illustrates the increasing amplitude of the
conductance oscillations near $eV\simeq \Delta _{0}.$ 
\begin{figure}[tbp]
\includegraphics[width=10cm,angle=0]{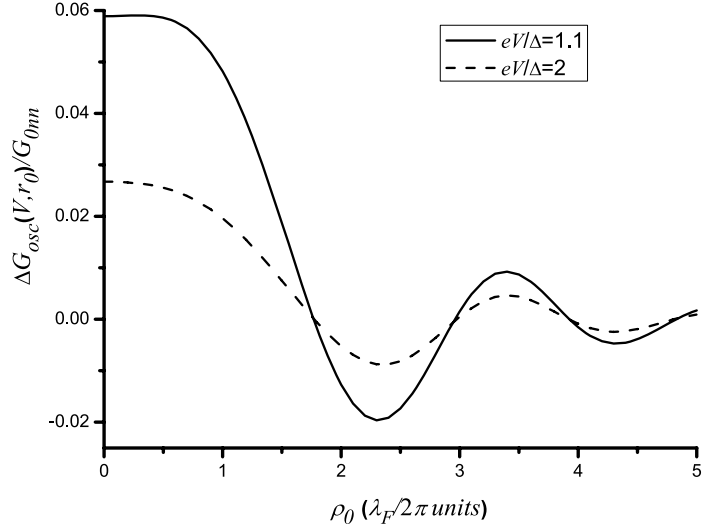}
\caption{Dependence of the normalized oscillatory part of the conductance $%
\Delta G_{osc}/G_{0ns}$, Eq.~(\protect\ref{dG_N}), on the distance $\protect%
\rho _{0}$ between the defect and the contact axis for two values of the
applied voltage. The defect is situated in the normal metal at a depth $%
z_{0}=5\lambdabar _{F}.$ The dimensionless constant of interaction is taken
as $\widetilde{g}=0.01.$}
\label{Fig-model}
\end{figure}

For the defect in the superconducting half-space the oscillatory part of the
conductance consists of two terms%
\begin{equation}
\Delta G_{osc}\left( V,r_{0}\right) =-G_{0ns}\left( V\right) \frac{12}{\pi }%
\widetilde{g}\left( \frac{\lambdabar }{r_{0}}\right) ^{2}\left(
k_{F}z_{0}\right) ^{2}\sum\limits_{\alpha =\pm }\psi _{\alpha }\left(
eV\right) j_{1}\left( k_{\alpha }r_{0}\right) y_{1}\left( k_{\alpha
}r_{0}\right) ,  \label{dG_S}
\end{equation}%
where%
\begin{equation}
\psi _{\pm }=\left\{ 
\begin{array}{c}
u_{0} \\ 
v_{0}%
\end{array}%
\right. ,\quad k_{\pm }=\frac{\sqrt{2m^{\ast }}}{\hbar }\left[ \varepsilon
_{F}\pm \sqrt{\left( eV\right) ^{2}-\Delta _{0}^{2}}\right] ^{1/2}.
\label{psi_k}
\end{equation}%
In Eqs.~(\ref{G0})-(\ref{dG_S}) we neglected all small terms of the order of 
$\Delta _{0}/\varepsilon _{F}$ and $eV/\varepsilon _{F}$ . Nevertheless we
kept the second term in square brackets in the formula for $k_{\pm }$ (see,
Eq.(\ref{psi_k})) because for a relatively large $r_{0},$ $(\sqrt{%
(eV)^{2}-\Delta _{0}^{2}}/\varepsilon _{F})(r_{0}/\lambdabar _{F})\simeq 1,$
the phase shift of the oscillations may be important. In Fig.~4 we show the
difference between the dependencies of the normalized oscillatory parts of
the conductance $\Delta G_{osc}/G_{0ns}$ on the distance $\rho _{0}$ for a
contact between normal metals $\left( \Delta _{0}=0\right) $ and for a NS
contact. An observable shift of the conductance oscillations results from
the voltage dependence of wave vectors $k_{\pm }$ (\ref{psi_k}). 
\begin{figure}[tbp]
\includegraphics[width=10cm,angle=0]{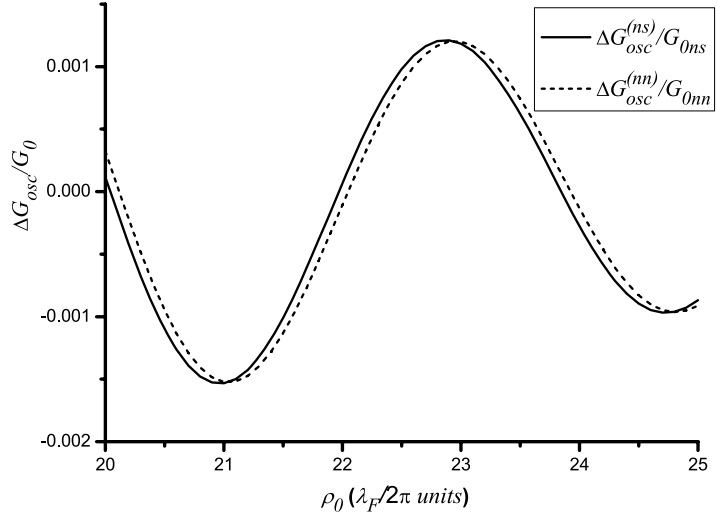}
\caption{The dependence of the oscillatory parts of the conductance $\Delta
G_{osc}/G_{0}$ (\protect\ref{dG_S}) on the distance $\protect\rho _{0}$
between the defect and contact axis for the contact between normal metals $%
(\Delta G_{osc}^{(nn)}/G_{0nn})$ and a NS contact $(\Delta
G_{osc}^{(ns)}/G_{0ns})$. The defect is situated in the right metal (the
superconductor) at a depth $10\lambdabar _{F}$; $eV/\Delta _{0}=5;$ $%
\widetilde{g}=0.01.$ }
\end{figure}

\section{Conclusion}

Thus, we have analyzed the conductance $G$ of a tunnel NS point contact with
a radius $a$ smaller than the Fermi wave length $\lambdabar _{F}$, at low
temperatures $\left( T=0\right) $ and for applied bias $eV$ larger than the
gap energy of the superconductor $\Delta _{0}.$ The effect of quantum
interference of quasiparticles scattered by a single defect situated in the
vicinity of the contact has been taken into account. We have shown that in
leading approximation in the parameters $eV/\varepsilon _{F}\ll 1,$ $\Delta
_{0}/\varepsilon _{F}\ll 1$ the conductance of a small NS contact is $%
G_{0ns}=G_{0nn}N_{s}\left( eV\right) $, Eq.~ (\ref{G0}), i.e., the product
of the conductance of the same contact between normal metals, $G_{0nn}\sim
a^{4}$, and the normalized density of states of the superconductor $%
N_{s}\left( eV\right) $, similar as for a planar tunnel contact. Although
such result is not unexpected and has been confirmed by experiment \cite{Pan}%
, for a contact of radius $a<$ $\lambdabar _{F}$ it was not obvious and it
is first obtained in this paper.

If the defect is situated in the normal metal the conductance displays
oscillations, the period of with is defined by the Fermi wave vector, $%
\Delta G_{osc}\left( V,r_{0}\right) \sim \sin 2k_{F}r_{0}$ at $k_{F}r_{0}\gg
1$ (Eq.~(\ref{dG_N}), Fig.~3), as for a contact between normal metals \cite%
{Avotina1}. In this case the defect plays the role of an additional
`barrier' between the normal and superconducting metals and results in
oscillations of the transmission coefficient. The underlying principle here
is similar to resonance transmission through a two-barrier system.

In the superconductor the electron wave incident on the contact from the
normal metal is transformed into a superposition of `electron-like' and
`hole-like' quasiparticles. In the case of location of the defect in the
superconducting half-space quantum interference takes place between partial
waves transmitted and those scattered by the defect, for both types of
quasiparticles independently (Eq.~(\ref{dG_S})). Although the difference
between wave vectors $k^{\left( \pm \right) }\left( eV\right) $ of
`electrons' and `holes' is small the shift $\left( k^{\left( +\right)
}-k^{\left( -\right) }\right) r_{0}$ between the two oscillations should be
observable (Fig.~4).

\section*{Appendix: Oscillations of the order parameter near the surface in
the presence of a defect.}

When calculating the conductance to first order in the transmission
probability we should know the order parameter $\Delta \left( \mathbf{r}
\right) $ in the limit of a nontransparent interface (surface), $%
U_{0}\rightarrow \infty $. According to Ref.~\cite{AGD}, 
\begin{equation}
\Delta ^{\ast }\left( \mathbf{r}\right) =\gamma T\sum\limits_{n=-\infty
}^{\infty }F_{\omega }^{+}\left( \mathbf{r},\mathbf{r}\right) \Theta \left(
\omega _{D}-\omega \right) ,  \tag{A1}  \label{deltaF}
\end{equation}%
where $\omega =\pi T\left( 2n+1\right) $ are the Matsubara frequencies. The
Fourier components $G_{\omega }\left( \mathbf{r},\mathbf{r}^{\prime }\right) 
$ and $F_{\omega }^{+}\left( \mathbf{r},\mathbf{r}\right) $ of Green's
functions satisfy the Gor'kov equations, which in the absence of a defect
potential have the form%
\begin{eqnarray}
\left( i\omega -\frac{\hbar ^{2}\nabla ^{2}}{2m^{\ast }}-\varepsilon
_{F}\right) G_{\omega }\left( \mathbf{r},\mathbf{r}^{\prime }\right) +\Delta
\left( \mathbf{r}\right) F_{\omega }^{+}\left( \mathbf{r},\mathbf{r}^{\prime
}\right) &=&\delta \left( \mathbf{r}-\mathbf{r}^{\prime }\right)  \TCItag{A2}
\label{Gor} \\
\left( i\omega +\frac{\hbar ^{2}\nabla ^{2}}{2m^{\ast }}+\varepsilon
_{F}\right) F_{\omega }^{+}\left( \mathbf{r},\mathbf{r}^{\prime }\right)
+\Delta ^{\ast }\left( \mathbf{r}\right) G_{\omega }\left( \mathbf{r},%
\mathbf{r}^{\prime }\right) &=&0.  \notag
\end{eqnarray}%
For a homogeneous superconductor $\Delta \left( \mathbf{r}\right) =\Delta
_{0}=\mathrm{const.}$ and the solutions $G_{\omega }\left( \mathbf{r},%
\mathbf{r}^{\prime }\right) =G_{\omega }^{\left( 0\right) }\left( \mathbf{r}-%
\mathbf{r}^{\prime }\right) $ and $F_{\omega }^{+}\left( \mathbf{r},\mathbf{r%
}^{\prime }\right) =F_{\omega }^{+\left( 0\right) }\left( \mathbf{r}-\mathbf{%
r}^{\prime }\right) $ of Eqs.(\ref{Gor}) can be found to be 
\begin{eqnarray}
G_{\omega }^{\left( 0\right) }\left( \mathbf{r}-\mathbf{r}^{\prime }\right)
&=&-\frac{\pi N\left( 0\right) }{k_{F}r}\left[ \cos k_{F}r+\frac{i\omega }{%
\sqrt{\Delta _{0}^{2}+\omega ^{2}}}\sin k_{F}r\right] \exp \left( -\frac{r}{%
v_{F}\hbar }\sqrt{\Delta _{0}^{2}+\omega ^{2}}\right) ,  \TCItag{A3}
\label{Gom} \\
F_{\omega }^{+\left( 0\right) }\left( \mathbf{r}-\mathbf{r}^{\prime }\right)
&=&\frac{\pi N\left( 0\right) \Delta _{0}^{\ast }}{\sqrt{\Delta
_{0}^{2}+\omega ^{2}}}\frac{\sin k_{F}r}{k_{F}r}\exp \left( -\frac{r}{%
v_{F}\hbar }\sqrt{\Delta _{0}^{2}+\omega ^{2}}\right) ,  \TCItag{A4}
\label{Fom}
\end{eqnarray}%
where $r=\left\vert \mathbf{r}-\mathbf{r}^{\prime }\right\vert ,$ $v_{F}$ is
the Fermi velocity, $\omega \ll \varepsilon _{F}.$ For the semi-infinite
superconducting half-space any component of the matrix Green function 
\begin{equation}
\widehat{G}_{\omega }^{\left( s\right) }\left( \mathbf{r},\mathbf{r}^{\prime
}\right) =\left( 
\begin{array}{cc}
G_{\omega }^{\left( s\right) }\left( \mathbf{r},\mathbf{r}^{\prime }\right)
& F_{\omega }^{\left( s\right) }\left( \mathbf{r},\mathbf{r}^{\prime }\right)
\\ 
F_{\omega }^{+\left( s\right) }\left( \mathbf{r},\mathbf{r}^{\prime }\right)
& -G_{-\omega }^{\left( s\right) }\left( \mathbf{r}^{\prime },\mathbf{r}%
\right)%
\end{array}%
\right)  \tag{A5}
\end{equation}%
can be written as%
\begin{equation}
\widehat{G}_{\omega }^{\left( s\right) }\left( \mathbf{r},\mathbf{r}^{\prime
}\right) =\widehat{G}_{\omega }^{\left( 0\right) }\left( \mathbf{r}-\mathbf{r%
}^{\prime }\right) -\widehat{G}_{\omega }^{\left( 0\right) }\left( \mathbf{r}%
-\widetilde{\mathbf{r}}^{\prime }\right) ,  \tag{A6}  \label{semi_inf}
\end{equation}%
where $\widetilde{\mathbf{r}}^{\prime }=\left( x^{\prime },y^{\prime
},-z^{\prime }\right) .$ Equation~(\ref{semi_inf}) is exact and it provides
the zero value of $\Delta \left( \mathbf{r}\right) $ at the surface $z=0.$
The fact that the order parameter vanishes at the nontransparent interface
can by seen from Eq.(\ref{delta}).

The Green's function for the superconducting half-space in the presence of
the point defect can be found from the Dyson equation 
\begin{equation}
\widehat{G}_{\omega }\left( \mathbf{r},\mathbf{r}^{\prime }\right) =\widehat{%
G}_{\omega }^{\left( s\right) }\left( \mathbf{r},\mathbf{r}^{\prime }\right)
+\int d\mathbf{r}^{\prime \prime }\widehat{G}_{\omega }^{\left( s\right)
}\left( \mathbf{r},\mathbf{r}^{\prime \prime }\right) D\left( \left\vert 
\mathbf{r}^{\prime \prime }\mathbf{-r}_{0}\right\vert \right) \tau _{3}%
\widehat{G}_{\omega }\left( \mathbf{r}^{\prime \prime },\mathbf{r}^{\prime
}\right) ,  \tag{A7}
\end{equation}%
where $\tau _{3}$ is the Pauli matrix. Making use of the small radius of the
defect potential in the first order approximation in the interaction
constant $g$ (\ref{g}) we obtain 
\begin{gather}
F_{\omega }^{+}\left( \mathbf{r},\mathbf{r}\right) =F_{\omega }^{+\left(
s\right) }\left( \mathbf{r},\mathbf{r}^{\prime }\right) +  \tag{A8}
\label{F} \\
g\left[ F_{\omega }^{+\left( s\right) }\left( \mathbf{r},\mathbf{r}%
_{0}\right) G_{\omega }^{\left( s\right) }\left( \mathbf{r}_{0},\mathbf{r}%
^{\prime }\right) +G_{-\omega }^{\left( s\right) }\left( \mathbf{r}_{0},%
\mathbf{r}\right) F_{\omega }^{+\left( s\right) }\left( \mathbf{r}_{0},%
\mathbf{r}^{\prime }\right) \right] .  \notag
\end{gather}%
As a first step for the self-consistent solution, the functions $G_{\omega
}^{\left( 0\right) }\left( \mathbf{r}-\mathbf{r}^{\prime }\right) $ (\ref%
{Gom}) and $F_{\omega }^{+\left( 0\right) }\left( \mathbf{r}-\mathbf{r}%
^{\prime }\right) $ (\ref{Fom}) may be used. At $T\rightarrow 0$ the
summation over Matsubara frequencies in Eq.(\ref{deltaF}) can be replaced by
an integration. Substituting the Eqs.(\ref{Gom}), (\ref{Fom}) into Eq. (\ref%
{semi_inf}) and using Eq.(\ref{F}) we find the space distribution of the
order parameter (\ref{deltaF}) in the next (after $\Delta =\Delta _{0}=%
\mathrm{const.}$) approximation. 
\begin{gather}
\Delta \left( \mathbf{r}\right) =\Delta _{0}\left\{ 1-\frac{\sin 2k_{F}z}{%
2k_{F}z}\ln ^{-1}\left( \frac{2\omega _{D}}{\Delta _{0}}\right) \mathcal{K}%
\left( \frac{2\pi z}{\xi _{0}};\frac{\omega _{D}}{\Delta _{0}}\right)
+\right.   \tag{A9}  \label{D(r)} \\
\frac{1}{4\pi }\widetilde{g}\ln ^{-1}\left( \frac{2\omega _{D}}{\Delta _{0}}%
\right) \left[ \frac{\sin 2k_{F}s_{0}}{2\left( k_{F}s_{0}\right) ^{2}}%
\mathcal{K}\left( \frac{2\pi s_{0}}{\xi _{0}};\frac{\omega _{D}}{\Delta _{0}}%
\right) +\frac{\sin 2k_{F}\widetilde{s}_{0}}{2\left( k_{F}\widetilde{s}%
_{0}\right) ^{2}}\mathcal{K}\left( \frac{2\pi \widetilde{s}_{0}}{\xi _{0}};%
\frac{\omega _{D}}{\Delta _{0}}\right) \right.   \notag \\
\left. \left. -\frac{\sin k_{F}\left( s_{0}+\widetilde{s}_{0}\right) }{%
k_{F}^{2}s_{0}\widetilde{s}_{0}}\mathcal{K}\left( \frac{\pi \left( s_{0}+%
\widetilde{s}_{0}\right) }{\xi _{0}};\frac{\omega _{D}}{\Delta _{0}}\right) %
\right] \right\} .  \notag
\end{gather}%
Here%
\begin{equation}
\mathcal{K}\left( a;b\right) =\int\limits_{0}^{\text{arsh}b}dte^{-a\text{ch}%
t},  \tag{A10}
\end{equation}%
$s_{0}=\left\vert \mathbf{r}-\mathbf{r}_{0}\right\vert ;$ $\widetilde{s}%
_{0}=\left\vert \mathbf{r}-\widetilde{\mathbf{r}}_{0}\right\vert ,$ and $\xi
_{0}=\hbar v_{F}/\pi \Delta _{0}$ is the coherence length. At $ab\gg 1,$ $%
\mathcal{K}\left( a;b\right) \simeq K_{0}\left( a\right) $, the modified
Bessel function \cite{AS}. The Eq.(\ref{D(r)}) is valid at distances from
the defect larger than the characteristic radius of the potential $D\left(
\left\vert \mathbf{r-r}_{0}\right\vert \right) .$ The correction to the
constant value of the order parameter $\Delta _{0}$ decreases at small
distances $r\ll \xi _{0}$ from the surface or the defect according to a
power law, and vanishes exponentially ($\sim e^{-2\pi r/\xi _{0}}$) at
larger distances $r\gg \xi _{0}.$ A grey-scale plot of $\Delta \left( 
\mathbf{r}\right) $ obtained by means of Eq.(\ref{D(r)}) is presented in
Fig.~2. In the plot we used an unrealistically large value of the constant $%
\widetilde{g}$ in order to show the influence on the order parameter the
defect and the surface in the same plot. For realistic values $\widetilde{g}%
\sim 0.01$ the spatial oscillations of $\Delta \left( \mathbf{r}\right) $
resulting from the scattering by the defect have a much smaller amplitude
than the second term in the braces of Eq.(\ref{D(r)}). The matching
procedure can be continued when we put $\Delta \left( \mathbf{r}\right) $ of
Eq.(\ref{D(r)}) into Gor'kov's equations (Eqs.(\ref{Gor})) or BdG equations (%
\ref{BdG}). Unfortunately, starting with this step the solutions may be
obtained only numerically.

\begin{acknowledgments}
One of us (Yu.K) would like to acknowledge useful discussions with A.N.
Omelyanchouk, E.V. Bezuglyi, and S.V. Kuplevakhsky. This research was
supported partly by the program "Nanosystems nanomaterials, and
nanotechnology" of National Academy of Sciences of Ukraine and Fundamental
Research State Fund of Ukraine (project F 25.2/122).
\end{acknowledgments}

\end{document}